\documentclass[11pt]{article} 

\usepackage[utf8]{inputenc}

\usepackage{geometry} 
\usepackage{graphicx} 

\DeclareGraphicsExtensions{.pdf,.png,.jpg,.eps}

\usepackage{booktabs} 
\usepackage{array} 
\usepackage{paralist} 
\usepackage{verbatim} 
\usepackage{subfig} 

\usepackage{fancyhdr} 
\usepackage{sectsty}
\allsectionsfont{\sffamily\mdseries\upshape} 

\usepackage{amsmath}
\usepackage{amsthm}
\usepackage{amsbsy}

\usepackage[colorlinks=true,urlcolor=blue,citecolor=black]{hyperref}

\usepackage[round]{natbib}
\title{Quantifying Artifacts over Time: Interval Estimation of a Poisson Distribution using the Jeffreys Prior}
\author{Stephen A. Collins-Elliott}
\date{}

\begin{document}

	\maketitle
	\frenchspacing
	
	\begin{abstract}
		This article presents a new method for estimating the amount of an artifact class in use at a given moment in the past from a random assemblage of archaeological finds. This method is based on the use of simulation, since an analytical solution is computationally impractical. Estimating the number of artifacts in use at any time $t$ is shown to follow a Poisson distribution, which allows for credible intervals to be established using the Jeffreys prior. This estimator works from minimal assumptions about the dating and duration of finds, as well as the intensity of collection, and is applied to coinage from four Roman-period sites excavated by the Roman Peasant Project (2009-2014). The result provides for an estimation of the abundance of material according to an interval of certainty. 
		
		Keywords: 
		estimation, statistics, simulation, coinage, Poisson distribution
	\end{abstract}

	\section{Introduction}
	
	This paper aims to answer the following question: given a random assemblage of datable artifacts from a location, how much of a specific artifact class was in use there at a given time in the past? The need to quantify the abundance of archaeological finds---ceramics, glass, animal bones, faunal remains, small finds---is of clear importance, and should be feasible for any past artifact that can be construed quantitatively. And yet,  the mere quantification of a class or type of artifact, whether recovered from stratigraphic deposits or surface contexts, is not the same as the quantification of that class in use at that location in a specific period. While quantification has tended to address intra-class concerns, such as methods to obviate fragmentation of ceramic and faunal assemblages, the question of how to address abundance over time remains under-theorized, and current attempts run the risk of abusing central definitions of probability, such as with the widespread approach of dividing finds over their date range and then summing the fractions per year. Moreover, the raw display of quantities, often in irregular period-intervals, can give a potentially misleading impression of the quantities of artifacts in use over time.  Accordingly, I would like in this paper to offer a mathematically sound solution to the problem of quantifying finds over time under minimal assumptions, which can incorporate a measure of uncertainty into the estimation.
	
	In moving from quantities of finds as archaeological materials to estimating their abundance in the past at any one moment, one must confront several well-known problems. First is loss, that is, how much was used at that location at that moment that has not been deposited there and thus is not part of the archaeological assemblage. Second is the related problem of post-depositional processes, which could have perturbed deposited strata and resulted in  finds being re-deposited in later layers as residual materials \citep{haselgrove_inference_1985,orton_sampling_2000}. Third, the precise start and end dates of an artifact's use at a site are often unknown, even though its date of production can be known in certain cases.  In other words,  each artifact's duration, how long it was in use at that site before it was deposited, is generally unknown. Finally, it is necessary to account for the intensity of finds collection through some metric, whether surface area, volume, or time, since more intensive collection will result in larger assemblages of material.
	
	These issues can be accommodated through formal, mathematical means. 
	To start by taking context into account, finds contained within secondary or tertiary contexts, often seen as detritus removed from its primary point of use, have value as a set of random observations.  Sporadic losses (and finds) of objects have been subject to a series of unknowable historical and depositional processes and are thus apt for informing stochastic models. In formal terms, an archaeological assemblage composed of such random material, $X$, is itself a pre-determined sample, only a partial record from the actual set of material culture in use, $\Omega$, whose makeup is unknown:
	\begin{equation*}
	X \subset \Omega.
	\end{equation*} 
	An assemblage of finds  $X$ satisfies the criterion of being ``independently and identically distributed'' (i.i.d.)~observations, since they are unique objects, as they are exchangeable in the order of their recovery: having found one artifact in the set before another does not change the overall composition of $X$. Let duration be predicated upon an average rate of discard or disuse, here called $\gamma$, which can be incorporated into the problem even if unknown. Similarly, let the intensity of collection at a locus $d$ be denoted $h_d$.

	Obtaining estimates that contain statistical information on uncertainty is crucial for accurately assessing trends in the scale and scope of operations in the ancient economy, and I wish here in this paper to focus on quantifying coinage in the Roman countryside as a case study. Whether or not coin use was a prevalent feature of Roman rural economies has long been a subject of contention  \citetext{\citealt{crawford_money_1970}; \citealt[20-22]{howgego_supply_1992}; \citealt[131]{hollander_money_2007}; \citealt[93-101]{kay_romes_2014}}. Patterns of coin use have also been of great interest for their potential to establish site-types from a bottom-up approach, and detect broader cultural or economic relationships or factors that lay behind the creation of the material record \citep{reece_site-finds_1995,lockyear_site_2000}. To a significant degree, however, coins are frequently recovered from perturbed contexts, such as topsoil, leaving the history of their use largely uncertain. This uncertainty has occasioned considerable dispute about the importance and utility of these random losses and finds of coins \citetext{\citealt[2-4]{howgego_supply_1992}; \citealt{blackburn_coinage_2011}; \citealt{ellis_re-evaluating_2017}}. Productive investigation of the use of quantified data borne of random losses has tended to be neutralized or inhibited by the onus to explain the life-history of the object, or by referencing the myriad tendencies and factors which impact coin loss and retrieval  \citep[110-113]{evans_coins_2013}. Some of these (like precision in dating) can be accommodated quantitatively, but first some further comment on these points is  necessary.
	
	Coinage has been treated as special or distinct from other classes of material, since coins are viewed as objects which have a higher value than others and are not considered to have been discarded through consumption or breakage \citep[195-197]{lockyear_dating_2012}. That said, many other artifact-classes face the same issues of retrieval, reuse, and recycling, such as ceramics \citep[in summary, see][319-352]{pena_roman_2007}. Observations from comparative situations can be used perhaps to provide some meaningful parameters to frame survival rates owing to differential factors and assess the question of representativeness. In one study of coin loss and retrieval in Melbourne, Australia, \citet{frazer_`random_2010} showed that losses of coins were representative of those in circulation with respect to their mintage dates across all denominations, and that loss and recovery rates were related not just to the denominational value of the coin, but the size of the coin as well. A positive relationship between size and denomination however does not hold across all monetary systems. For example, higher-value Roman \textit{denarii}  are smaller than lower-value \textit{asses}. Cultural factors too have an impact beyond the economic value and dimensions of the coin. To take another modern example, \citet{griffiths_attrition_2002} found that US commemorative quarters issued from 1999 onward had a much higher attrition rate from circulation than the regular eagle quarters, a trend likely owed to the habit of coin-collecting. 
	
	The quantification of coinage in light of the issues of loss and recovery cannot be separated from the cultural dynamics of the society which used those coins, and emphasis on the broader cultural significance of coinage illustrates the range of contexts and uses which coinage held beyond its purpose as a form of money \citep{aarts_coins_2005,haselgrove_reassessing_2005,rowan_slipping_2009,kemmers_rethinking_2011}. Counts of coins as well in their archaeological context can also serve as indicators of other types of behavior aside from attesting to exchange, as for example in the analysis of depositional patterns at Pompeii \citep{ellis_re-evaluating_2017}. Such factors bear on the issue of quantification only in so far as establishing a formal relationship  between $X$, the obtained sample of finds, and $\Omega$, the unknown population of material culture in use at that place in the past. Theoretically, if the same social and cultural habits around retrieval and reuse, as well as roughly similar site-formation processes, have held broadly over the period of interest, then the same tendencies will affect all estimates uniformly. The set $X$ can then be taken to be representative of $\Omega$, just as the size of $X$ can be considered to be proportional to the size of $\Omega$, even though it may not be known absolutely. These assumptions are made for the purposes of the solution presented in this paper to the problem of quantification over time, yet they can and should certainly be explored in further depth.
	
	It is also necessary to clarify the aims of what the estimation of the frequency of coin use reflects. Monetization and circulation, two terms that are sometimes used synonymously with amounts of coinage recovered at archaeological sites, should not be equated outright with the abundance of coinage, in light of their technical, economic definitions. ``Monetization''---defined as the act of rendering things into monetary terms or the production of currency by an issuing authority \citetext{\citealt[243-244]{katsari_monetization_2008}; \citealt[18-19]{von_reden_money_2010}}---and ``circulation''---defined as the total amount of currency issued minus that which has hitherto been withdrawn by that authority---speak more to aspects of the political economy based on the perspective of the minting authority. The frequency or abundance of coinage in use in the past can, and should, be treated as something distinct, but nonetheless significant for evaluating the degree to which currency-based exchange took place within a given spatial or social context. 
	
\begin{figure}[t!] 
		\includegraphics[width=1\textwidth]{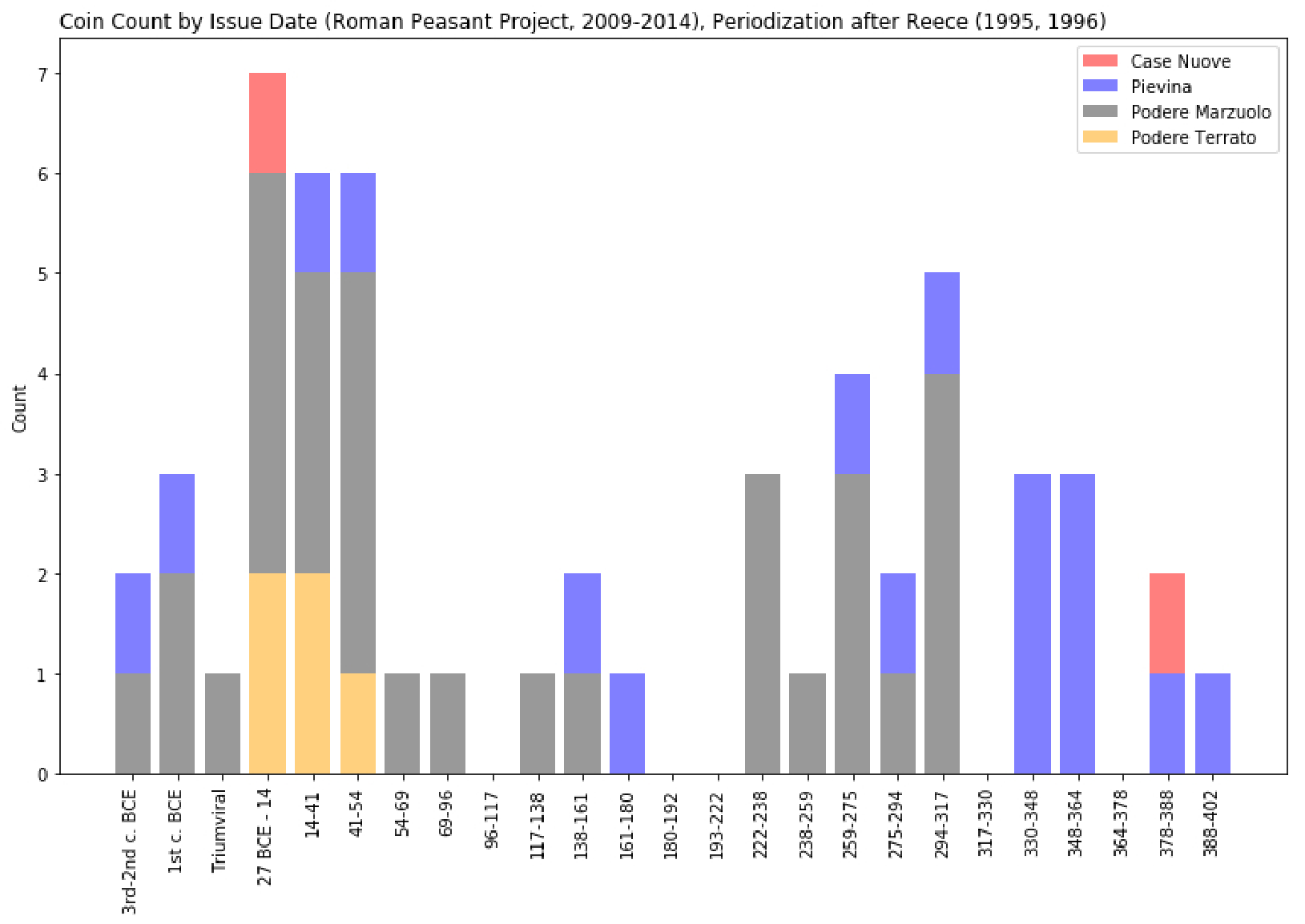}
		\caption[]{Quantification of coins by mintage date from RPP sites, according to the periodization of \citet{reece_site-finds_1995,reece_interpretation_1996}. \label{fig:fig1}}
	\end{figure}

	Quantifying random coins has tended to proceed using raw counts by mintage date, grouped per phase or period, or relative to a mean or as a percentage of a total assemblage  \citep{reece_use_1984,reece_site-finds_1995,reece_interpretation_1996,lockyear_site_2000,hoyer_overview_2018}. By way of illustration, we can take as a case study the coinage from the  Roman Peasant Project (RPP), which excavated multiple rural sites in central Tuscany from 2009 to 2014  \citep{ghisleni_excavating_2011,vaccaro_excavating_2013}. Four sites in particular, Pievina, Case Nuove, Podere Terrato, and Podere Marzuolo, yielded datable finds of coins from secondary or tertiary contexts (the site of Tombarelle only produced one illegible coin). Arraying the coins by their mint date using the periodization developed by \citet{reece_site-finds_1995} shows two general periods of roughly equivalent increases in the amounts of coinage: ca. early first century CE and ca. late third century CE (Fig. \ref{fig:fig1}). Yet, this chart leaves out coins which cannot be dated precisely to a period, but which can nevertheless be attributed to a broader interval of time, such as late Antique fractional issues that are nearly illegible. Moreover, the mint dates do not speak to when the coins were used at the site: they could have come into use and could have been deposited significantly later than their date of production. This is certainly the case with well-worn second- and early first-century BCE coins at Podere Marzuolo. Finally, the sites were excavated at different intensities, which is not accounted for in this chart: Podere Marzuolo was excavated for two months and Podere Terrato in less than one, leaving one in doubt as to whether higher amounts of coinage are reflective of more thorough investigations. 
	
	I wish to proceed first by discussing previous approaches to the problem of quantifying the abundance of artifact-types over time, called ``aoristic analysis.'' The mainstay technique of this approach is simply to divide artifacts by their potential time intervals and then add those fractions in each interval. As I wish to argue, this method is invalid under the accepted mathematical definition of probability. It is necessary instead to define a formula which calculates the expected value of material at each moment $t$. Thus, I provide an equation to estimate the abundance of a material over time when the precise durations of artifacts' use are unknown. The enormity of the problem means that it must be solved by simulation, rather than direct calculation. Using this approach, I estimate the relative amounts of coinage which could have been in use over time, from the Roman Peasant Project sites discussed above, incorporating the aforementioned variables of duration ($\gamma$), and fieldwork intensity ($h$).

	\section{Background}

	There has been much recent work on modeling the frequency or abundance of an artifact class over time  under the name of ``aoristic analysis,'' a label originating in the field of geography \citep{ratcliffe_aoristic_2000}. In brief, aoristic analysis seeks to quantify an event or occurrence when the precise time of the event is unknown. As such, the topic has obvious applications in archaeology, and has been of increasing interest the more that computational methods are applied in the field \citep{crema_probabilistic_2010,crema_aoristic_2011,crema_modelling_2012,bevan_measuring_2013,crema_time_2015}. Essentially, the procedure entails dividing a quantity of material over its possible time periods, and then summing its fractions per unit of time: that is, if one has two artifacts which could have been in use over a span of 100 years, their quantity would be $2/100$ in any year. This technique has been employed in more or less formal terms  well before it became called ``aoristic,'' a label applied first by \citet{johnson_aoristic_2004}  in the case of archaeology. In a recent review of aoristic analysis, \citet{baxter_reinventing_2016} note the widespread use of this technique and its organic development in the field: the earliest detectable traces of an aoristic analysis can be found in \citet{carlson_computer_1983} and \citet{fentress_counting_1988}, the latter of whom pioneered the method for quantifying African Red Slip ceramics over time. Currently, the practice remains diffuse and is still described by a variety of names, such as the use of ``weighted averages'' in Roman ceramics \citep[e.g.][]{fentress_accounting_2004,di_giuseppe_black-gloss_2012}.
	
	While work into ``aoristic'' values shows the interest in and need to develop ways to quantify material over time, their validity is subject to question. The major problem with aoristic values is that they appear to assume an additive property which does not accord with that established for the mathematical definition of probability as a chance event from 0 to 1.  Rather, aoristic values misconstrue the probability that an object occurs in a given year or moment as a fraction of its quantity, to be summed or averaged. Indeed, \citet[126]{baxter_reinventing_2016} already raised the question of what exactly aoristic sums represent, but did not venture to build a more sound mathematical footing for estimating the frequency of artifacts in use over time. 
	
	\begin{figure}[t!] 
		\includegraphics[width=1\textwidth]{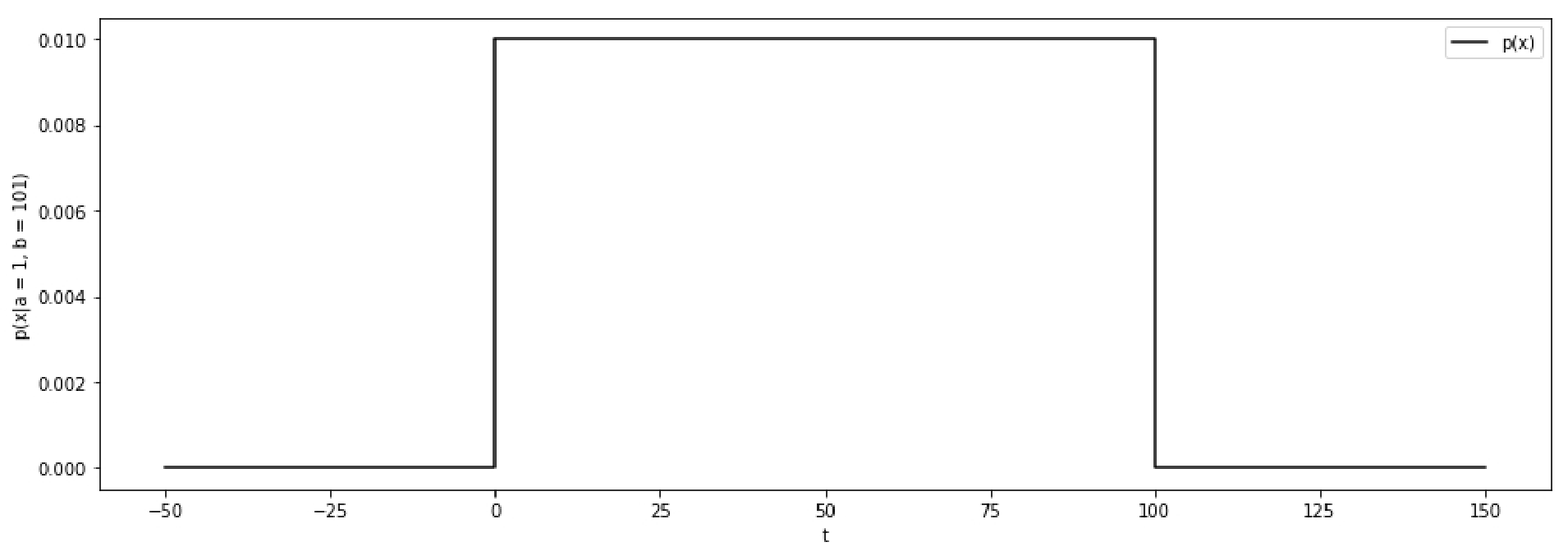}
		\caption[]{The uniform distribution $\text{U}(a,b)$, expressing an equal probability of the chance occurrence of an artifact occurring from year 1 to 101 CE.\label{fig:uniform}}
	\end{figure}
	
	To give more detail, the use of probability theory in artifact dating was exemplified by  \citet[100]{buck_bayesian_1996}, who applied the uniform distribution in modeling the probability that an artifact was used at any one time between moments $a$ and $b$. The probability of this event, $p(x)$, will be $1/(b-a)$. For example, if $a$ is the year 1 CE, and $b$ is the year 101 CE, then the probability that that object was in use at any one year in the first century CE will be $1/100$ (Fig. \ref{fig:uniform}). Yet, where aoristic analysis appears to go astray comes in dealing with multiple artifacts in light of treating their occurrence as events. If one had two different artifacts, $x_1$ and $x_2$, the probability that those two artifacts were in use in the same year, as independent events, is $p(x_1) \cdot p(x_2)$, which will be $1/10000$ (not $2/100$). A probability does not quantify the abundance of a material, but rather expresses the chance of its occurrence. Aoristic analysis makes what appears to be a reasonable leap in equating the probability of something as a fraction of its quantity, but this is unfortunately invalid under the definition of probability as the chance of an event.

	Accordingly, I wish to offer a new approach, to define an estimator of abundance over time which is mathematically valid and feasible to compute. Furthermore, it is desirable to keep it simple, to reduce the amount of inputs to the most minimal variables with the fewest amount of assumptions: the earliest possible date of the artifact ($a$), the latest possible date of the artifact ($b$), and an average rate of use and discard of the material ($\gamma$), which is based on the probability that the artifact-type either continued in use or went out of use. The intensity of collection ($h$) must be accounted for, as well. As the following section will show, given the amount of uncertainty within the possible actual dates of use and possible combinations thereof,  simulation is necessary as a means of computation---direct calculation is not practical.

	\section{Methodology}

	To start with a simple example, it is clear that if one coin had one year in which it could have been in use, the abundance of coinage would be 1 for that year. If there are two years in which it could have been used, then there are three possibilities, that there is one coin in use in the first year but not the second, that there is one coin in use in the second year but not in the first, or that the coin is in use in both years. Each of these can be designated as three separate potential events, respectively denoted $x_1, x_2$, and $x_3$:

	\begin{table}[h!]
		\centering
		\begin{tabular}{ll|ll|ll}
			\toprule
			\multicolumn{2}{c}{$x_1$} & \multicolumn{2}{c}{$x_2$} & \multicolumn{2}{c}{$x_3$} \\
			\midrule
			1 & 0 & 0 & 1 & 1 & 1\\
			\bottomrule
		\end{tabular}
	\end{table}

	If there are two coins, then there are nine different possible combinations of their chronological pattern: 
	
	\pagebreak
	
	\begin{table}[h!]
		\centering
		\begin{tabular}{lll|ll|ll | ll|ll|ll | ll|ll|ll}
			\toprule
			event &		\multicolumn{2}{c}{$x_1$} & \multicolumn{2}{c}{$x_2$} & \multicolumn{2}{c}{$x_3$} & 		\multicolumn{2}{c}{$x_4$} & \multicolumn{2}{c}{$x_5$} & \multicolumn{2}{c}{$x_6$} & 		\multicolumn{2}{c}{$x_7$} & \multicolumn{2}{c}{$x_8$} & \multicolumn{2}{c}{$x_9$} \\
			\midrule
			first coin&	 1 & 0 & 0 & 1 & 1 & 1  &  1 & 0 & 0 & 1 & 1 & 1   & 1 & 0 & 0 & 1 & 1 & 1\\
			second coin&	 1 & 0 & 1 & 0 & 1 & 0  &  0 & 1 & 0 & 1 & 0 & 1   & 1 & 1 & 1 & 1 & 1 & 1\\
			\midrule
			total coins &2 & 0 & 1 & 1 & 2 & 1  & 1 & 1 & 0 & 2 & 1 & 2  & 	2 & 1 & 1 & 2 & 2 & 2\\
			\bottomrule
		\end{tabular}
	\end{table}

	Each one of these events has an equal chance of having occurred, and can accordingly be expressed as a probability, $p(x)$, which  is here equal to $1/9$. To find the amount of coinage in either year, it is just a matter of calculating the expected value. Like the mean value of the roll of a six-sided die, which will be 3.5 ($\textrm{E} [X]=1\cdot {\frac {1}{6}}+2\cdot {\frac {1}{6}}+3\cdot {\frac {1}{6}}+4\cdot {\frac {1}{6}}+5\cdot {\frac {1}{6}}+6\cdot {\frac {1}{6}}=3.5$), the expected value of the coinage in any one year can be found by summing the amount of coins in any one year multiplied its probability. For example, the first year will have the expected value of:
	\begin{equation*}
	\textrm{E} [X] = 2 \cdot \frac{1}{9} + 1 \cdot \frac{1}{9} + 2 \cdot \frac{1}{9} + 1 \cdot \frac{1}{9} + 0 \cdot \frac{1}{9} + 1 \cdot \frac{1}{9} + 2 \cdot \frac{1}{9} + 1 \cdot \frac{1}{9} + 2 \cdot \frac{1}{9} = \frac{4}{3}
	\end{equation*}
	which will also be the same expected value of amount of coins in that second year.
	
	It should be clear that increasing the number of coins in the assemblage, as well as the number of years in the potential duration, will make the computation of each of these combinations overly time-consuming, since, if $n$ is the number of artifacts, and $t$ is the number of years in which that coin could have been in use, then the number of possible combinations of intervals is
	\begin{equation*}
	C_I = \left(\sum_{l = 1}^{t} l \right)^n.
	\end{equation*}
	Just five coins over one hundred years would result in $(100 + 99 + 98 + \cdots + 1)^5$, or  $5050^5$ (3.28 E+18), combinations of potential durations of use. Combined with the fact that in an actual implementation of the method every coin will have its own particular date range, calculating the expected value for every possible combination (or any other interval of time) is impracticable. Nevertheless, a formal, mathematical approach is necessary to solve the problem, and can be undertaken using random simulation.
	
	I wish to describe the technique of simulation in plain terms. I take an assemblage of coins, interpreted as random finds, under the assumptions that they have an earliest possible date of their issue (\textit{terminus post quem}) and a latest possible date of their loss (\textit{terminus ante quem}). I then undertake a series of repeated iterations. In each iteration, I randomly allot a duration of use to each coin within its potential date range, which is predicated upon the average rate of discard of the artifact class ($\gamma$). I then calculate the number of coins in use at every moment in time, record that amount, and then perform another iteration, for $k$ number of iterations. The expected value of coins can then be calculated, where each simulated run represents one potential event, to arrive at a point estimate (like $4/3$ above in the case of two coins in two years). Thus, it is possible to have an approximate estimate of the expected value $E[X]$, even if its actual computation is not feasible. This routine can be summarized in the following algorithm:
	\begin{enumerate}
		\item Choose a number of iterations ($k$) to calculate the expected value through simulation.
		\item Start an iteration to simulate one potential series of dates of artifacts of a sample $X$.
		\item Generate a random set of date ranges for each artifact in that sample, according to their own particular dates of use ($a,b$) and the duration rate ($\gamma$).
		\item Store that list of simulated date ranges. Go back to step 2, and repeat for the total number of iterations ($k$).
		\item Find an approximate value $E[X]$ for each year $t$ by taking the total set of simulated date ranges, summing the amount of simulated objects in use, and then dividing by the total number of simulations.
	\end{enumerate}
	

	\subsection{Formal Definition}\label{definitions}
	
	This section outlines in formal terms the steps described above. Let  $X = \{x_1, \ldots, x_n\}$ represent a set of artifacts that was recovered from a site $d$. In this model, the process or intention behind their deposition cannot be accounted for, and  they are taken to be random losses and finds of an artifact type.  Every $x_i$ will have a date of issue, $a_i$, and a latest possible date of final disuse, loss, or deposition,~$b_i$.
	
	As noted above, it is not realistic to calculate $E[X]$ by analytical means, working through every possible combination of potential periods of use, and so to estimate it at a moment of time $t$, it is necessary to employ a process of simulation. Each iteration of the simulation will create  a random, potential interval of use for each $x_i$ within the range from $a_i$ to $b_i$. Let this potential interval be denoted by the start and end dates $\alpha_i$ and $\beta_i$, such that 
	\begin{equation*}
	a_i < \alpha_i < \beta_i \leq b_i.
	\end{equation*}
	If $b_i$ represents the instant of the abandonment of the site or the end of a phase at the site, the selection of a $\beta_i < b_i$ would mean that the artifact was lost or deposited before the end of the site or phase, while  $\beta_i = b_i$  would mean that the artifact  went out of use with the end of the phase or site. 
	
	The additional variable of the rate of discard ($\gamma$) can be taken into account as follows. While  $\alpha_{i,j}$ is randomly chosen between its date of issue ($a$) and date of disuse ($b$), the random variable $\beta_{i,j}$ is modeled on a Geometric distribution, since it can be viewed as a series of Bernoulli trials: for each year $t$ (the ``trial''), whether the artifact remains in use (event $B_0$, the ``failure''), or whether it falls out of use (event $B_1$, the ``success''), is dependent upon a rate $\gamma$. For the sake of this model, if the artifact is lost and recovered within the same year, it is considered to remain in use. For example, if any one artifact is considered to have an even chance of continuing in use or of being lost in a year, $\gamma = 1/2$. In this case, the variable of $\beta_{i,j}$ is  selected  at random,   not uniformly like $\alpha_{i,j}$, but as according to the probability expressed by the Geometric distribution, since it is contingent upon the occurrence of all previous events:
	$B_1$, the event of disuse, and the product of all $B_0$ after moment $\alpha$ that it  continued to remain in use, for $t-\alpha \in \{1,2,3,\dots\}$:
	\begin{equation*}\label{lossrate}
	p(\beta = t) =  (1-\gamma)^{t-\alpha-1}\gamma
	\end{equation*}
	The rate of $\gamma = 1/2$, though giving even chances that the object continues in circulation or is discarded that year, may not be the most uninformative, nor even the most realistic value to choose. In certain cases, expert opinion or observation of artifact-use derived from analogous situations may be able to provide a use rate which is realistic to the material class of the artifact. In other cases, the additional sampling of Monte Carlo values of a range of possible rates might be useful (if there is anywhere between a 1\% and 10\% rate of discard, $0.01 \leq \gamma \leq 0.10$). Or, if treating an object or event which is regarded as instantaneous, $\gamma = 1$.
	
	The simulation is run for $k$ iterations. Let $j$ be any one  iterated run of the simulation, and so for $j = 1, \ldots, k$,   $v_{i,j} = (\alpha_{i,j}, \beta_{i,j})$ represents any one pair of random simulated dates of use for an artifact $x_i \in X_d = \{x_1, \ldots, x_n\}$ from a locus $d$. As described above, let  $\alpha_{i,j} \sim \text{U}(a_i,b_i)$ be randomly chosen, and $\beta_{i,j} \sim \text{Geom}(\gamma)$ from moment $t-\alpha$. Let $\mathbf{v_j} = [v_{1,j}, \ldots,  v_{n,j}]$ represent the list of all pairs of simulated dates for any one simulation run. There will thus be  $n \cdot k$ number of simulated dates of  $\mathbf{v_1}, \ldots, \mathbf{v_k}$.

	In order to calculate the expected value of artifacts in use for a year $t$, let $u_{i,j}$ represent a moment in which that artifact is in use within a simulated date range, such that $\alpha_{i,j} \leq u_{i,j} \leq \beta_{i,j}$. Using the notation of the Iverson bracket, let $[t = u_{i,j}]$ indicate a value of 1 if that year $t$ falls within the simulated date range, and 0 in all other cases. Thus, the sum of all artifacts for any year $t$ within any one set of simulated date ranges $\mathbf{v_j}$ can be denoted:
	\begin{equation}
	w_j(t) = \sum_{i = 1}^{n}  [t = u_{i,j}]
	\end{equation}
	Let the expected value $E[X]$ for any given year $t$ be denoted $U(t)$. It will thus consist of the mean value of all $w_1(t),\ldots,w_k(t)$:
	\begin{equation}\label{usefunction}
	U(t) =  \frac{1}{k} \sum_{j = 1}^{k}    w_j(t)
	\end{equation}
	with variance
	\begin{equation}
	\sigma^2 = \frac{1}{k} \sum_{j = 1}^{k} [w_j(t) - U(t)]^2.
	\end{equation}

	Finally, since more thorough investigation will yield  more finds,  the result must be weighted by the intensity of fieldwork, $h$, whether expressed in volume of soil, surface area, or time, of that locus $d$. The final estimation will therefore represent abundance of artifacts in use at a given moment in time as a density, during the site's occupation (here, $m^2$ is used as the metric of $h$, and so the estimation will be artifacts per m$^2$ at time $t$). The abundance of artifacts within a locus $d$ at time $t$ can thus be defined as:
	\begin{equation}\label{estimator}
	\Upsilon_d (t) = \frac{1}{h_d} {U}(t).
	\end{equation}

	\begin{figure}[t!] 
		\includegraphics[width=1\textwidth]{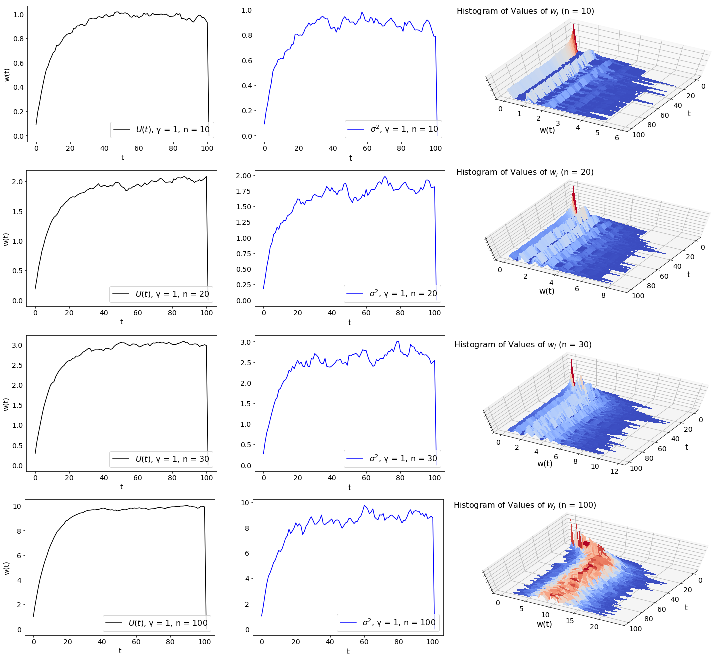}
		\caption[]{From top to bottom, simulated values of $w_j$, for samples of size $n = 10, 20, 30,$ and $100$, where every artifact has start date $a = 1$ and end date $b = 101$, and with $\gamma = 0.1$. The plot of $U(t)$ is to the left, $\sigma^2$ in the center, with the histogram of all values over $t$ to right.  \label{fig:histograms}}
	\end{figure}

	\subsection{The Poisson Distribution}
	
\begin{figure}[t!] 
		\includegraphics[width=1\textwidth]{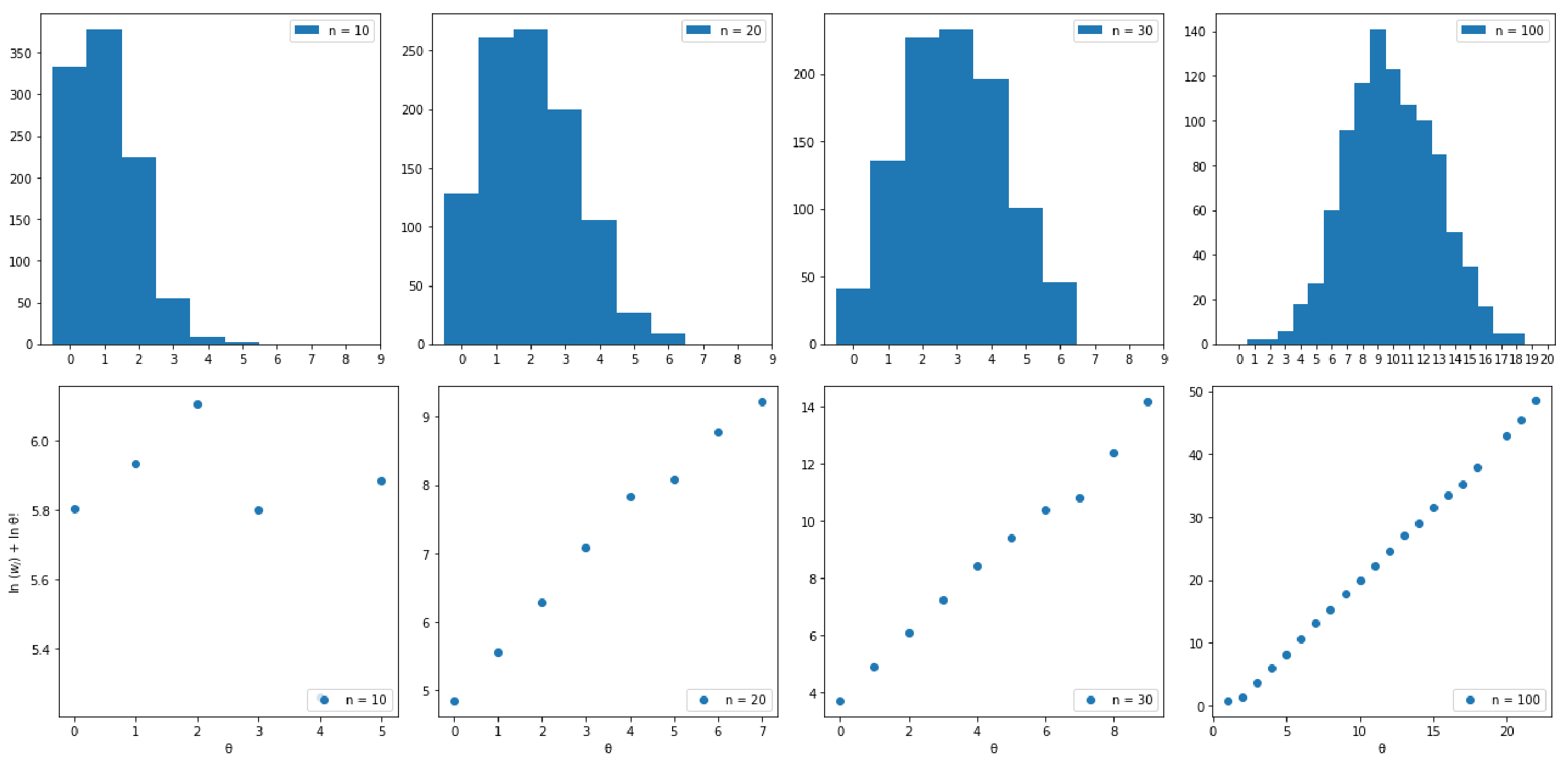}
		\caption[]{Above: histograms of the simulated values of $w_j$ when $t = 50$, for samples of size $n = 10, 20, 30,$ and $100$, where every artifact has start date $a = 1$ and end date $b = 101$, and with $\gamma = 0.1$. Below: evaluation of Poissonness \citep{hoaglin_poissonness_1980}, to determine the goodness of fit for the Poisson distribution.\label{fig:poissonness}}
	\end{figure}

	Simulating the number of artifacts each year constitutes an act of estimating discrete events. Moreover, initial examination of histograms and an apparent correlation between $U(t)$ and $\sigma^2$ suggested that the estimation of artifacts in each year $t$ might be modeled using a Poisson distribution (Fig. \ref{fig:histograms}). The Poisson distribution is a rate-based probability distribution which deals with discrete, independent occurrences, using a rate $\lambda$ for a number of occurrences or arrivals, $\theta$. In other words, the rate parameter $\lambda$ represents the average number of occurrences (i.e., the expected value). The probability mass function, which expresses the probability of observing $\theta$ number of artifacts (whether $\theta = 0, 1, 2, 3, \ldots$), is of the form
	\begin{equation*}
	p(\theta) = \frac {\lambda ^{\theta}e^{-\lambda }}{\theta!}.
	\end{equation*}
	Both the mean and the variance of the Poisson distribution are equal to $\lambda$. In order to assess whether or not the quantification of artifacts in use at year $t$ can be modeled on the Poisson distribution, I used the Poissonness plot to evaluate the distribution of the simulated values of $w_j(t)$ \citetext{\citealt{hoaglin_poissonness_1980}; \citealt[348-358]{hoaglin_exploring_1985}}. To apply the formula of \citet[146, Eq. 2.2]{hoaglin_poissonness_1980} to the notation of this paper, where the observation consists of $w_j(t)$ and $\lambda = U(t)$, the log of the above equation can be rewritten as:
	\begin{equation*}
	\ln (w_j) = \ln (k) - U(t) + \theta \ln(U(t)) - \ln(\theta!).
	\end{equation*}
	The procedure then is to plot $\ln (w_j) + \ln (\theta!)$ against $\theta$. If the graphic representation of the data falls in an approximately straight line, then the data meet well with  the Poisson distribution. Plotting trial data of different sample sizes revealed that the the Poisson distribution closely fit the simulations of $w_j(t)$ for samples of and around $n > 20$, with a poorer fit for samples less than that (Fig. \ref{fig:poissonness}). That said, given that $U(t) \approx \sigma^2$ even for small samples, the poor characterization of Poissonness at small sample sizes should not be taken to obviate the general pattern observed, especially in light of the theoretical motivations of the use of the distribution to evaluate the arrival of discrete events as the observation of simulated finds.

	The use of the Poisson distribution matters for the way in which a credible interval (CI) can be established around $U(t)$. Since the mean and variance are identical for the Poisson distribution (both are $\lambda$), one could apply a 95\% CI at $\pm   1.96 \sqrt{\lambda}$ around the mean, but this  holds for the Poisson distribution only where it has a large sample size, since for large $n$ it starts to approach the form of a normal distribution. The same CI will not be useful when the sample size is small and the Poission distribution does not tend toward normality. Rather, it should be noted that a number of methods have been developed to find intervals for the Poisson distribution \citep{patil_comparison_2012}. Here, a Bayesian CI which uses the Jeffreys prior is preferable, and can be computed for a Poisson distribution analytically, as discussed by \citet[27-28]{brown_interval_2003} and \citet[251-254]{nadarajah_bayesian_2015}. For a CI according to $100 \times (1 - \zeta)$ \% certainty, based on an uncertainty measure $\zeta$ \citep[259-262]{bernardo_bayesian_1994}, the lower ($\lambda_L$) and upper ($\lambda_U$) bounds of the interval estimate can be found using the following expression \citep[28, Eq. 15]{brown_interval_2003}, here adapted to the variables defined in the paper:
	\begin{equation*}
	\lambda_L \sim \text{Gamma}  \left(\frac{\zeta}{2},  \sum_{j=1}^{k} w_j + \frac{1}{2}, k\right) , \lambda_U \sim \text{Gamma} \left({1 - \frac{\zeta}{2}}, \sum_{j=1}^{k} w_j + \frac{1}{2}, k\right)
	\end{equation*}
	Thus, the estimation of quantities of artifacts over time can be calculated using a Bayesian credible interval. This has the advantage of being straightforward to compute, and,  if other, more informative prior probabilities are known about the quantities of artifacts in use per year in a given case, that prior knowledge can be accommodated into the posterior interval estimate through a Gamma-Poisson model of inference.

	\section{Implementation: Frequency of Coin Use through Random Losses from the Roman Peasant Project}\label{sec:rpp}

	In this section, I examine how implementing the above method of estimation provides a richer picture of quantification over time than that of counts per issue date (Fig.~1). Table~\ref{sites} shows four rural sites excavated by the Roman Peasant Project (2009-2014), which yielded assemblages of non-modern, datable coinage from  stratified and topsoil contexts, secondary or tertiary deposits (i.e., non-hoard, non-intentional burials of coins) at the sites of Pievina (2009), Case Nuove (2010), Podere Terrato (2011), and Podere Marzuolo (2012-2013). The intensity of fieldwork at each site, $h_d$, is measured in the excavated surface area of each site. Let the $b$ value of the coins be determined by the end of each phase of a site (Podere Terrato is the only site with a single phase), and $a$ be determined by the issue date of the coin or the start of occupation of the site, whichever is later. The intensity of fieldwork ($h_d$) was taken into account as the final step, after the calculation of the credible interval.
	\begin{table}	
		\centering
		\begin{tabular}{l c c r }
			\hline
			Site ($d$) & Area Excavated ($h$) & Period $(a,b)$ & $n$\\
			\hline
			Pievina (PI) & 764.63 m$^2$ & (-200,470) &  34\\
			Case Nuove (CN) & 995.29 m$^2$ &(-30,450) & 9 \\
			Podere Terrato (PT) &636.65 m$^2$ & (20,60) & 5 \\
			Podere Marzuolo (MZ)&550.49 m$^2$ &(-30,320) & 31\\
			\hline
		\end{tabular}
		\caption{Overview of sites, excavated by the Roman Peasant Project. The number of coins $n$ are all datable ancient issues (i.e., modern and illegible coins are excluded).\label{sites}}
	\end{table}
	
\begin{figure}[p!] 
		\includegraphics[width=1\textwidth]{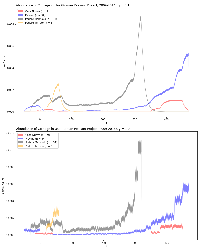}
		\caption[]{The abundance of coins in use at any one year $\Upsilon_d(t)$ quantified from the finds collected by the Roman Peasant Project, with a 90\% CI and under the assumptions of the discard rate at $\gamma = 0.10$ (above) and $\gamma = 1.0$ (below). \label{fig:rpp1}}
	\end{figure}

	Changing the rate parameter alters the quantities estimated of coins per $m^2$ according to the sample $X$, but given that $X$ is taken to be proportional to $\Omega$, this does not affect the inter-site comparison of the results too much. Estimation was done first using a discard rate of $\gamma = 0.10$ and then using a discard rate of $\gamma = 1$, both with a 90\% credible interval (Fig. \ref{fig:rpp1}). A discard rate of 10\% means that 10\% of the coinage, once entering into use on a site, would be discarded or lost, while a 100\% discard rate means that it has a guaranteed loss in the subsequent year. In contrast to the initial graphic display of the quantities of coins per mintage date (Fig. \ref{fig:fig1}), this graph contains information on the potential duration of use of each coin  and the effect which the intensity of collection has had on the quantities of coins recovered. It also presents the estimation according to a measure of certainty. Incorporating this information into the quantification of coin use over time shows where an initial autopsy of the raw data might be misleading. Although Pievina has a slightly higher amount of coinage, the duration of its occupation and its intensity of excavation means that it does not possess overall higher measures of abundance of use as Podere Marzuolo. Similarly, Podere Terrato, with only five coins dating to the Julio-Claudian period, shows a fairly high density for its short period of occupation, in contrast to the other sites.
	
	Since one can assume that there were more coins in circulation at each site than those that had been recovered (not every coin used on a site will have been lost there), the plot can be interpreted \textit{stricto sensu} as the minimum threshold for the abundance of coin use. If the amount over time given the sample $X$ is proportional to the amount in $\Omega$, consistently across all sites (a frequent assumption in the quantification of archaeological materials), then this plot shows when each site undergoes higher and lower levels of coin use relative to one another. Accordingly, Podere Marzuolo (MZ) underwent a period of steadily increasing coin use throughout the later third century CE, whereas Pievina saw a similar trend in the fourth and fifth century CE. Both of these trends were greater than those visible in the late first century BCE and first century CE. To be sure, differing intensities of coin use are owed to a host of complex factors, not just monetization and money supply, but demographic factors, such as intensity of occupation and human mobility, as well as the cultural norms of exchange (whether the payment of debts is negotiated through cash or credit) also play a role. Toward developing arguments about those factors, however, the ability to provide a probabilistic estimate of the quantity of coinage, as an artifact, in use over time is essential, in contrast to potentially erroneous impressions about the quantity of surviving material based on raw counts alone absent of a spatial and chronological framework.
	
	\section{Conclusions}
	
	This paper has sought to provide a mathematically valid way to quantify material over time under minimal assumptions, using coinage as an example. Due to the number of uncertain variables in the problem, simulation was used to randomly allot artifacts to time intervals within a \textit{terminus post quem} and \textit{terminus ante quem}, and then to find the mean of those simulated values. It was shown that the estimation of artifact abundance in a year $t$ follows a Poisson distribution well for reasonably sized samples. Moreover, small sample sizes nevertheless elicited results where mean and variance are related, suggesting that the use of the Poisson distribution in low sample sizes was not inappropriate, which altogether accords with the theoretical motivations behind the use of the Poisson distribution for this situation. Calculating a credible interval using the Jeffreys prior also provided upper and lower bounds to the estimation according to a fixed rate of discard $\gamma$. 
	
	Different discard rates ($\gamma$) can be selected to accommodate the situation at hand. A value of $\gamma = 1$ ensures that the object will only be in use for that one year, and can be used to model the abundance of instantaneous events of other archaeological phenomena (for example, shipwrecks), rather than of durations of use. The selection of $\gamma$, however, need not rely on a single arbitrary value. Rather, $\gamma$ can be specified with in a range, for example, to stipulate a rate of discard between 1\% and 5\%  ($0.01 \leq \gamma \leq 0.05$). Or, the selection of $\gamma$ may be predicated upon a probability distribution, which dictates whether there are more likely values than others. Experimentation with these variable rates shows that the Poissonness of the estimation will be affected, or rather, that the result will be a complex mixture of Poisson distributions, such that other methods of calculating a credible interval  should be explored instead, like a highest density interval (HDI).
	
	In sum, the use of probability densities and interval estimation means that uncertainty about the abundance of material in use over time can be transfered to more complex models of ancient social and economic behavior. Future work can benefit from simulation as a means to deal with uncertainty borne of archaeological data, such as rates of coin use and loss. Adapting the method presented here to other datasets that are mainstays of evidence, whether ceramics, counts of sites in a region and their duration of occupation, or instances of shipwrecks, will enable more accurate measurement of the magnitude and extent of their occurrence.

	\section*{Acknowledgments}
	
	I would like to thank David Stone, Aleydis Van de Moortel, and Jack Hanson for their helpful comments on this paper, as well as Myles Lavan for alerting me to \citet{carlson_computer_1983}. Thanks also to my colleague Vasileios Maroulas at the University of Tennessee for the opportunity to present this research at the Mathematical Data Science Seminar and for the questions and comments from the seminar participants. This work was made possible thanks to a University of Tennessee Professional Development Award. An earlier version of this paper was presented at the 117th Annual Meeting of the Archaeological Institute of America, San Francisco, CA. The coins from the 2009-2010 seasons were identified by Flavia Marani, with those from the 2011-2014 seasons by the author.  I wish to thank the project directors, Kim Bowes, Cam Grey, Marielena Ghisleni, and Emanuele Vaccaro, for their permission to study the coinage from their project. Any errors are my own.

	\section*{Appendix: Scripts and Data}
	
	The scripts used in this paper were written in Python using \texttt{NumPy} and \texttt{SciPy} packages \citep{jones_scipy:_2001,oliphant_guide_2015}, \texttt{rpy2} \citep{gauthier_rpy2_2014}, and  \texttt{matplotlib} \citep{hunter_matplotlib:_2007}. The script and necessary data from the Roman Peasant Project are attached as supplementary files, and are also available at \url{http://www.github.com/scollinselliott/lostchange}.
	\begin{itemize}
		\item \textsf{estimator.py} contains the basic algorithm to quantify the number of artifacts in use per year, as well as to produce the graphics in this paper.
		\item Two csv files \textsf{input-rpp.csv} and \textsf{outputcontext-rpp.csv} contain the data from the Roman Peasant Project used in Section \ref{sec:rpp}. The basic form of input used by the Python script uses a csv file with the following columns: ``Site'' is used to denote the locus $d$, and can be used to refer to a region, site, or phase. ``Domain'' refers to the material of the artifact-class (in case working from a table with multiple artifact-types). ``Count'' refers to the quantities of that artifact contained in that row (here, 1, for each coin). ``Date1'' and ``Date2'' refer to $a$ and $b$ respectively. ``SFid'' refers to the small find inventory number of the coin.
	\end{itemize}

\bibliography{myrefsbibtex}


 \end{document}